\documentclass[preprint,showpacs,preprintnumbers,amsmath,amssymb,12 pt.,one-column,double-spaced]{revtex4}

\usepackage{graphicx}
\usepackage{dcolumn}
\usepackage{bm}
\usepackage{color}

\begin{document}

\preprint{APS/123-QED}

\title{Electric Transport of a Single Crystal Iron Chalcogenide FeSe Superconductor: Evidence of Symmetry Breakdown Nematicity and Additional Ultrafast Dirac Cone-Like Carriers}

\author{K. K. Huynh$^1$}

\author{Y. Tanabe$^2$}
 \thanks{Corresponding author: youichi@sspns.phys.tohoku.ac.jp}

\author{T. Urata$^2$}

\author{H. Oguro$^3$}

\author{S. Heguri$^1$}

\author{K. Watanabe$^3$}

\author{K. Tanigaki$^{1, 2}$}
\thanks{Corresponding author: tanigaki@sspns.phys.tohoku.ac.jp}

\affiliation{$^1$WPI-Advanced Institutes of Materials Research, Tohoku University, Aoba, Aramaki, Aoba-ku, Sendai, 980-8578, Japan}

\affiliation{$^2$Department of Physics, Graduate School of Science, Tohoku University, Aoba, Aramaki, Aoba-ku, Sendai, 980-8578, Japan}

\affiliation{$^3$High Field Laboratory for Superconducting Materials, Institute for Materials Research, Tohoku University, Sendai 980-8577, Japan}

\date{\today}

\begin{abstract}
  An SDW antiferromagnetic (SDW-AF) low temperature phase transition is generally observed and the AF spin fluctuations are considered to play an important role for the superconductivity paring mechanism in FeAs superconductors.
  However, a similar magnetic phase transition is not observed in FeSe superconductors, which has caused considerable discussion. 
  We report on the intrinsic electronic states of FeSe as elucidated by electric transport measurements under magnetic fields using a high quality single crystal. 
  A mobility spectrum analysis, an ab initio method that does not make assumptions on the transport parameters in a multicarrier system, provides very important and clear evidence that another hidden order, most likely the symmetry broken from the tetragonal C4 symmetry to the C2 symmetry nematicity associated with the selective d-orbital splitting, exists in the case of superconducting FeSe other than the AF magnetic order spin fluctuations. 
  The intrinsic low temperature phase in FeSe is in the almost compensated semimetallic states but is additionally accompanied by Dirac cone like ultrafast electrons $\sim$ 10$^4$cm$^2$(VS)$^{-1}$ as minority carriers. 
\end{abstract}

\pacs{74.70.Xa, 74.25.Dw, 72.15.Gd, 75.47.-m}

\maketitle

\section{Introduction}
After the discovery of the high temperature superconductivity of LaFeAsO$_{1-x}$F$_x$ by Kamihara $et$ $al$. in 2008, a variety of superconducting compounds such as FePn (Pn = P, As) and FeCh (Ch = Se, Te) were found and studied \cite{1, 2, 3, 4, 5, 6, 7}.
FeSe has the simplest crystal structure among the iron-based superconducting families and is composed of two-dimensional FeSe blocks stacked along the c-axis direction \cite{8, 9}, and therefore has become a very important platform for understanding the mechanism of superconductivity as well as the intriguing structural / electronic phase transitions that occur when temperature is lowered.
In the FeAs superconducting families, a magnetic SDW phase transition generally occurs together with a structural phase transition from tetragonal to orthorhombic.
It is generally considered that spin fluctuations of the low temperature antiferromagnetic (AFM) phase play an important role for mediating superconductivity.
This is generally denoted as the S$\pm$ mechanism \cite{10}.
As another candidate for the superconductivity mechanism, the orbital fluctuation in a d-multiband system has been proposed \cite{11}.
FeSe has been proven to show that no magnetic phase transitions are observed even though a kink in electric conductivity as a function of temperature is observed \cite{12}.
Therefore, a surge of interest has arisen to compare the FeAs and FeSe superconducting families, and hitherto a lot of experiments have been performed so far.
However, due to the difficulty in achieving single crystal growth, there have been few experimental reports on the electronic states of FeSe \cite{13, 14}.

An interesting report regarding the superconductivity has recently been made for FeSe by Xue $et$ $al$., showing that a single layer FeSe epitaxially grown on a SrTiO$_3$ substrate showed a $T_{\rm c}$ much higher than what has been ever seen in the Fe superconducting family \cite{15}.
However, due to the difficulty in elucidating the intrinsic electronic states of a single layer of FeSe, the real nature of superconductivity in FeSe is still an important open question.
Recently, a method of high quality single crystal growth of 1:1 stoichiometric FeSe has been reported \cite{16}.
Therefore, it is very important and timely to study the electric transport properties of FeSe single crystals.

In this letter, we report on the detailed electronic structure of a FeSe bulk single crystal studied by transport measurements under magnetic fields.
The experimental data, by employing ab initio mobility spectrum analysis without making assumptions on the carrier numbers  successfully applied to Ba(FeAs)$_{2}$ in our previous study \cite{20}, clearly demonstrate that the electron and the hole pockets with almost equivalent carrier numbers of $\sim$10$^{20}$ cm$^{-3}$ are present in the majority of carrier bands, which is consistent with a carrier-compensated semimetallic feature.
In addition, we intriguingly find that another minority band with the carrier number of $\sim$10$^{18}$ cm$^{-3}$ is also present. 
Surprisingly, these minority electron carriers show an ultrafast carrier mobility of $\sim$ 10$^4$cm$^2$(VS)$^{-1}$.
Moreover, both a remarkable reduction in carrier number and an enhancement in carrier mobility were simultaneously observed below 120 K higher than the structural transition temperature.
Although no antiferromagnetic (AFM) ordering has been reported in pure FeSe \cite{12}, the electronic structure of orthorhombic FeSe single crystal shows a clear change in the electronic structure at a temperature apart from the structural transition. 
The present results provide a very important clue to understand the mechanism of superconductivity and the electronic phase transition of iron-based superconductors.

\section{Experiment}
FeSe single crystal was grown by a vapor transport method using an Fe$_{1.1}$Se and KCl/AlCl$_3$ mixture with the molar ratio of 1 : 0.5 \cite{16}.
The Fe$_{1.1}$Se and KCl/AlCl$_3$ mixture was prepared inside an Ar globe box and sealed into a quartz tube under the He gas pressure of 10$^{-2}$ Pa. 
The quartz ampoule was placed in a tube furnace with thermal gradient.
In the present case, the position of raw materials was kept at 390 $^{\circ}$C. 
After 40 - 50 days, thin flakes of FeSe single crystals ($\sim$ 500 $\times$ 500 $\times$ 30 $\mu$m) were obtained at a lower temperature position inside the quartz ampoule. 
During the crystal growth process, the temperature at the lower end of the quartz ampoule was kept higher than the melting point of KCl/AlCl$_3$.

Figure 1(a) shows a typical FeSe single crystal grown from the method described above.
The crystal exhibited clear \textit{ab}-plane.  The crystal was cut rectangular in shape for measurements of electric transport properties.
X-ray diffraction spectrum of the crystal, measured on the \textit{ab}-plane using a SmartLab 9MTP (RIGAKU) diffractometer, displayed sharp $(00\mathrm{l})$ reflections arising from the tetragonal phase of FeSe \cite{8, 9}.
In measurements of transport properties, electrodes were made on the ab-plane using silver paste (figure 1(a)).
Magnetic field was applied along the c-axis of the sample during the measurements of Hall-resistance and magnetoresistance \cite{18}.
Figure 1(c) shows the temperature ($T$) dependence of electrical resistivity for the FeSe single crystal.
The superconducting transition temperature ($T_{\rm c}$) was defined as the end point of the superconducting transition in the $\rho$ - T curve.
In order to estimate the ratio-of-residual-resistance (RRR), the resistivity at $0\,\mathrm{K}$ ($\rho(0 \mathrm{K})$) was extrapolated from the $\rho(T)$ curve in the temperature range of $12\,\mathrm{K} \leq T \leq 60\,\mathrm{K}$.
The obtained value of RRR = $\rho$(300 K)/$\rho$(0 K) was 138, and significantly higher than that of polycrystalline samples (RRR $\sim$ 5) \cite{8}.

\section{Results}
Figure 1(c) shows a typical temperature dependence of the resistivity ($\rho$($T$)) observed in our FeSe single crystals. 
A clear kink in $\rho$($T$) is always observed at the structural transition $T^{\ast}$ = 90 K, hinting that an electronic structure transition takes place.
Interestingly, the change in electronic states indicated by the kink became clear when we carried out the conductivity experiments under a magnetic field ($B$) (magnetoresistance ($R_{xx}$($B$)/$R$(0)) and the Hall resistivity (R$_H$) above and below $T^{\ast}$, as shown in figures 1(d) and 1(e).
Below the $T^{\ast}$ = 90 K, $R_{xx}$($B$)/$R$(0) increased with a decrease in $T$ and became 300 $\%$ at 12 K under 9 T, while it was below 1 $\%$ above $T^{\ast}$.
The gradient of $\rho_{yx}$ becomes maximum at 80 K and the non-linear $\rho_{yx}$ was developed with a decrease in $T$ \cite{17}.
At 12 K, a clear sign change in $\rho_{yx}$ was confirmed.
These abrupt changes in the $B$-dependence of the magneto-transport properties unambiguously suggest a drastic reconstruction of the Fermi surfaces across the transition at $T^{\ast}$. 
From the view point of the semiclassical theory of transport, such a large enhancement of magnetoresistance can be directly linked to a big jump in the carrier mobility, whereas a sign change of R$_H$ can appear only in the case where a large change in mobility takes place between electron-like and hole-like carriers. 
As we shall see in the analysis below, these curious behaviors of the transport properties are indeed associated with a very unique evolution of the electronic structure as a function of temperature.

\section{Analyses of the magnetotransport properties}
Because the current material is a d-multiband semimetal, it is generally difficult to have a clear interpretation of the magneto-transport data due to the mixing of many possible Fermi pockets.
A conventional approach to this situation is to construct a model by hypothesizing the number of carrier types, where each is characterized by two parameters: carrier density and mobility.
These can be estimated by applying the least-square fitting technique to the $B$-dependences of R$_{\rm H}$ and $R_{xx}$($B$)/$R$(0).
The results obtained in such an analysis, however, is highly dependent on the preliminary assumption about the number of pockets.
In order to avoid any ambiguities for making reasonable interpretations of the transport properties of FeSe single crystal in the whole temperature range, we have employed two different analysis methodologies in this paper.
At first, the so-called mobility ($\mu$-) spectrum analysis was applied to deduce the intrinsic band picture from magneto-transport at the lowest temperature.
As will be shown below, the results of the $\mu$-spectrum analysis clearly indicate the compensation between electron-like and hole-like carriers in FeSe.
This observation allows us to used a simple compensated two-band model to analyze the magnetotransport at higher temperatures.

\subsection{$\mu$-spectrum analysis}
Recently, we have successfully applied the $\mu$-spectrum analysis for inprepretation of the transport data of  Ba(FeAs)$_2$ \cite{20}, and the same methodology is employed in the current paper.
In the following part, we give a brief description of the method.

At first, the notations of normalized longitudinal and transverse magnetoconductivities were introduced:
\begin{eqnarray}
  X(B) = \frac{1}{\sigma_0}\frac{\rho_{xx}(B)}{\rho_{xx}^2(B) + \rho_{yx}^2(B)} \,;\label{eq:X}\\
  Y(B) = \frac{1}{\sigma_0}\frac{\rho_{yx}(B)}{\rho_{xx}^2(B) + \rho_{yx}^2(B)} \,.\label{eq:Y}.
\end{eqnarray}
Here $\sigma_0$ is the conductivity at zero magnetic field.
The normalized conductivities $X(B)$ and $Y(B)$ calculated from the present experimental data are shown in Fig. 2.
Instead of making assumptions on the number of Fermi pockets, we evaluate the distribution of the carrier numbers in the mobility spectrum space \cite{18, 19, 20}.
In the description of $\mu$-spectrum analysis, the transverse and the longitudinal conductivities under $B$ can be described by the distributions $s^n$ and $s^p$:
\begin{eqnarray}
X(B)=\int_{0}^{\infty} \frac{s^n(\mu)}{1+\mu^2B^2} + \int_{0}^{\infty} \frac{s^p(\mu)}{1+\mu^2B^2} = X^n(B) +X^p(B) \label{eq:X_mu-spec},\\
Y(B)=\int_{0}^{\infty} \frac{\mu s^n(\mu)}{1+\mu^2B^2} + \int_{0}^{\infty} \frac{\mu s^p(\mu)}{1+\mu^2B^2} = Y^n(B) +Y^p(B) \label{eq:Y_mu-spec}.
\end{eqnarray} 
Here $n$ and $p$ represent the electron and the hole carriers, respectively and $\mu$ is the carrier mobility. 
A set of [$X^k$($B$), $Y^k$($B$)] denotes the partial longitudinal and transverse conductivities for electron  ($k$ = $n$) or hole ($k$ = $p$) carriers. 

The Kronig-Kramer (KK) transformation applied to equations \eqref{eq:X_mu-spec} and \eqref{eq:Y_mu-spec} allows one to separate the conduction of electron-like carriers from those of hole-like ones as following:
\begin{eqnarray}
  \frac{1}{\pi}\mathcal{P}\int_{-\infty}^{+\infty}\frac{dB^\prime}{B - B^\prime}X(B) &= Y^{(p)}(B) - Y^{(n)}(B)\,, \label{eq:KK-transform_a}\\
    \frac{1}{\pi}\mathcal{P}\int_{-\infty}^{+\infty}\frac{dB^\prime}{B - B^\prime}Y(B) &= -X^{(p)}(B) + X^{(n)}(B)\,.\label{eq:KK-transform_b}.
\end{eqnarray}
Finally, using equations \eqref{eq:X_mu-spec}, \eqref{eq:Y_mu-spec}, \eqref{eq:KK-transform_a}, and \eqref{eq:KK-transform_b}, the partial conductivities [$X^k$($B$), $Y^k$($B$)] of electron-like and hole-like carriers can be calculated. 
For applying the Kronig-Kramer transformation (\eqref{eq:KK-transform_a} and \eqref{eq:KK-transform_b}) to the experimental data in the finite range of $B$, the analytic representation of the experimental data $X(B)$ and $Y(B)$ is evaluated by fitting the real data of $X(B)$ and $Y(B)$ to the linear combinations of Lorentzian components \cite{18, maxima}:
\begin{eqnarray}
  \label{eq:replacement_a}
  X^\prime(B) = \sum_i{\frac{\alpha_i}{1 + {\mu_{\alpha,i}}^2B^2}}\,, \\
  \label{eq:replacement_b}
  Y^\prime(B) = \sum_i{\frac{\beta_iB}{1 + {\mu_{\beta,i}}^2B^2}}\,.
\end{eqnarray}
The finalized parameters of the Lorentzian terms are listed in \ref{tab:lorentzian-components}.
It is clear in Fig. 2 that the two kinds of datasets are almost identical to each other.
The KK transformations were performed on the analytic representations .
The partial conductivities for electron-like and hole-like carriers, $[X^{(n)}(B),\,Y^{(n)}(B)]$ and $[X^{(p)}(B),\,Y^{(p)}(B)]$, obtained from the calculations are shown as the blue and the orange curves, respectively.

\begin{table*}
  \caption{Lorentzian components.}
  \begin{tabular}{c c c | c c c}
    \hline
    \multicolumn{3}{c|}{$X$} & \multicolumn{3}{c}{$Y$} \tabularnewline
    No.($i$) & $\mu_{\alpha,i}$ ($\text{m}^2(\text{Vs})^{-1}$) & $\alpha_i$ &  
    No.($i$) & $\mu_{\beta,i}$ ($\text{m}^2(\text{Vs})^{-1}$) & $\beta_i$ \tabularnewline
    \hline
    $1$ & $0.691$ & $0.206$ &  $1$ & $0.38$ & $-55.129$
    \tabularnewline
    $2$ & $0.068$ & $0.135$ &  $2$ & $0.378$ & $52.86$
    \tabularnewline
    $3$ & $0.6854$ & $0.166$ &  $3$ & $0.483$ & $3.826$
    \tabularnewline
    $4$ &  &  &  $4$ & $0.567$ & $-1.662$
    \tabularnewline
    \hline
  \end{tabular}  
  \label{tab:lorentzian-components}
\end{table*}

In a logarithmic equally-spaced grid of the $\mu$-space, the normalized conductivities $X^{(k)}(B)$ and $Y^{(k)}(B)$ in \eqref{eq:X} and \eqref{eq:Y} can be approximated as followed:

\begin{eqnarray}
  \label{eq:mu-spectrum-cal-X}
  X^{(k)}(B) 
    &= \sum^N_{i=0} \frac{1}{1 + \exp(2(m_i + b))} \times e^{m_i} s^{(k)}(m_i) \Delta m \nonumber \\
    &=  \sum^N_{i=0} \frac{1}{1 + \exp(2(m_i + b))} \times h_i\,;\\
  \label{eq:mu-spectrum-cal-Y}
  |Y^{(k)}(B)| 
    &= \sum^N_{i=0} \frac{1}{2 \cosh(m_i + b)} \times e^{m_i} s^{(k)}(m_i) \Delta m \nonumber \\
    &=  \sum^N_{i=0} \frac{1}{2 \cosh(m_i + b)} \times h_i \,,
\end{eqnarray}
where $\mu = e^m$, $B = e^b$, and $h_i = e^{m_i} s^{(k)}(m_i) \Delta m$.
Here $N$ is the total number of points used in the approximation and $\Delta m$ is the distance between two $m_i$ points.
In order to estimate the $\mu$-spectra of FeSe single crystal, models including 100 points were generated and the models were then independently fitted to the datasets using the program fityk \cite{Fityk}. 
For both $k = n$ and $k = p$, the spectrum extracted from $X^k(B)$ is identical with that obtained from $Y^k(B)$, confirming the validity of our analyses.
The $\mu$-spectra $s^k$($\mu$) were successfully evaluated for electron-like and hole-like carriers using the partial conductivities [$X^n$($B$), $Y^n$($B$)] and [$X^p$($B$), $Y^p$($B$)] as shown in Fig. 3.
A nearly single-peak structure centered at around  $\mu$ $\sim$ 1000 cm$^2$(VS)$^{-1}$ with the carrier number $P$ $\sim$ 10$^{20}$cm$^{-3}$ was found in the hole-like carrier region.
In contrast, a somewhat broad, double peak structure was deduced in the electron-like carrier region. 
The value of $\sigma$$_0$$s^n$($\mu$)/e$\mu$ showed that the 1st peak is at around $\mu$ $\sim$ 1000 cm$^2$(VS)$^{-1}$ with the carrier number $N$ $\sim$ 9 $\times$ 10$^{19}$cm$^{-3}$.
This estimated carrier concentration is comparable with $\sigma_0s^p(\mu)$/e$\mu$ for hole like carriers in the mobility spectrum space, which is indicative of a semimetallic feature of the FeAs single crystal (the first peak deduced from the transport analysis in the electron region can be assigned to the main electron-like carriers).
Taking these experimental facts into account, the stoichiometric FeSe single crystal is a carrier compensated semimetal and an almost equivalent amount of carriers with similar mobilities exist in both hole and electron regions.
Intriguingly, a broad peak structure (the second peak) was observed with the carrier number of $N$ $\sim$ 10$^{18}$ cm$^{-3}$ in the electron region.
Since high mobility carriers could be dominant in $X$($B$) and $Y$($B$) in the semiclassical transport theory, these minority electron-like carriers with $\mu$ $\sim$ 10000 cm$^2$(VS)$^{-1}$ should play a significant role in the electrical transport of FeSe at low $B$'s even when its carrier number is much smaller than that of the majority electron carriers.

\subsection{Two-band analysis at higher temperatures}
An important result of the $\mu$-spectrum is the confirmation of the compensation in carrier number between the hole-like and the electron-like pockets in the material.
Moreover, density functional theory (DFT) calculations have shown that the tetragonal phase of FeSe exhibits a semimetal-like FS \cite{21}.
These allow us to extend our analysis to higher temperatures in the framework of a semimetallic approximation, i.e. $N$ = $P$, and consequently study the changes of the electronic structures in terms of the transport parameters.
At high $T$'s, since $R_{xx}$($B$)/$R$(0) and the non-linear $\rho_{yx}$ were significantly suppressed, the mobility spectrum analysis could not monitor the whole electronic structure under the normal magnetic fields of the present experiments. 
In a two-carrier type semiclassical approximation in the low-$B$ limit, the zero-field resistivity($\rho_{xx}$ (0)), $R_{xx}$($B$)/$R$(0), and $\rho_{yx}$ were described as
\begin{eqnarray}
 \rho_{xx}(0) &=& \frac{1}{e(n_e\mu_e + n_h\mu_h)} \label{3},\\
 R_{xx}(B))/R(0) &=& \frac{n_en_h\mu_e\mu_h(\mu_e + \mu_h)^2B^2}{(n_e\mu_e + n_h\mu_h)^2} \label{4},\\
 \rho_{yx} &=& \frac{(-n_e\mu_e^2 + n_h\mu_h^2)B}{e(n_e\mu_e + n_h\mu_h)^2}. \label{5}.
\end{eqnarray}
Here $n_e$ and $n_h$ are the carrier numbers and $\mu_h$ and $\mu_e$ are the mobilities of electrons and holes.
Employing Eqs. (11) - (13) under the condition of $n_e$ = $n_h$, the transport parameters could be evaluated analytically.
The deduced carrier numbers and their mobilities of $N$ = $P$, $\mu_e$ and $\mu_h$ are displayed in Figs. 4 (a) and (b).
Below 120 K, $n$ gradually decreased with a decrease in $T$ and dropped below $T^{\ast}$, and at low-$T$'s $n$ became $\sim$ 9 $\times$ 10$^{19}$cm$^{-3}$.
Compared with the carrier numbers ($n_{MS}$$^k$, $k$ = $n$, $p$) estimated from the mobility spectrum analysis, $n_{MS}$$^n$ $\sim$ 9 $\times$ 10$^{19}$cm$^{-3}$ and $n_{MS}$$^p$ $\sim$ 1 $\times$ 10$^{20}$cm$^{-3}$ are reasonable.
Therefore, 80 - 90 $\%$ of the carriers were killed at low-$T$'s.
Both $\mu_e$ and $\mu_h$ gradually increased with a decrease in $T$ below 120 K.

\section{Discussions}
From the mobility spectrum analysis, the low temperature electronic structure of FeSe was successfully investigated.
The hole-like carriers can be explained in terms of almost uniform single-peak mobilities in the mobility spectrum space.
On the other hand, the electron-like carriers could be divided in two components in the mobility space.
The first peak with low mobility of $\mu$ $\sim$ 1000 cm$^2$(VS)$^{-1}$ is almost compensated with that of the hole-like carriers.
The second peak with the much higher mobility of $\mu$ $\sim$ 10000 cm$^2$(VS)$^{-1}$ plays a key role in the electrical transport under low $B$.
In the collinear-type AFM phase at low $T$'s in the parent compounds of iron-based superconducting families, the electronic bands composed of both minority Dirac cone carriers with high mobility and majority normal carriers accommodated in the almost compensated parabolic bands were reported \cite{20, 22, 23, 24}. 
Because no AFM phase was reported in FeSe under ambient pressure conditions \cite{12}, the evolution of the electronic states as a function of $T$ in FeSe is indeed intriguing when compared with the other iron based superconducting families.

In the low temperature orthogonal phase, the electronic structure of FeSe is represented as an almost compensated semimetal state with ultrafast minority electron-like carriers.
Moreover, since the reduction in the carrier number as well as the enhancement in the carrier mobility was confirmed below around 120 K (importantly higher than the structural phase transition temperature being indicative of the occurrence of another hidden order), the electronic structure of FeSe should unambiguously change at low $T$'s.
The band calculations predicted that the electron and the hole compensation takes place in the high temperature tetragonal phase with carrier numbers of $N$, $P$ = 2.91$\times$10$^{21}$ cm$^{-3}$ as well as in the low temperature collinear-type AFM phase with the carrier number of $N$ = 2.7$\times$10$^{20}$ cm$^{-3}$ ($P$ = 1.8$\times$10$^{20}$ cm$^{-3}$) \cite{25}. 
However, no AFM phase was reported in the low temperature orthogonal phase of FeSe.
Recent ARPES experiments in single crystal FeSe clearly displayed the energy band splitting associated with the orbitals selectively involved in the band at low temperatures, which is consistent with the interpretation of the electronically driven nematicity \cite{26}.
In this case, the lift-off of the energy bands could reduce not only the carrier number but also the electron-electron scattering in the electron and the hole FS's.
Consequently, both suppression in number and enhancement in mobility of carriers starting at a certain temperature above $T^{\ast}$ could reasonably be understood in terms of the electronically nematic ordering.
In Fe based superconductors, the pairing mechanism of the superconductivity and the origin of the electronic nematicity have been discussed in terms of magnetic and orbital fluctuations \cite{10, 11, 27}.
Our present results indicate that the intrinsic scenario for the superconducting mechanism of FeSe-based superconductors involves the mediation via the pure orbital fluctuations, which is in strong contrast with other FeAs superconductors.

Finally, we would like to discuss the origin of the ultrafast minority electron-like carriers observed in the $\mu$-spectrum.
Such high $\mu$ carriers can originate from the enhancement of the anisotropic FS's due to orbital splitting.
Since $\mu$ is directly proportional to the Fermi velocity and the effective mass, this anisotropy may exert a strong influence on the k-position of FS's, giving a broad distribution of the mobility spectrum.
Another scenario is that a Dirac cone is created due to the splitting of energy bands.
In the latter case, the orbital splitting of energy band at low $T$'s can lift up a crossing between the two bands with distinctly different orbital characters ($d_{xy}$ and $d_{yz}$) in the vicinity of the Fermi energy \cite{28}.
Since the Dirac cones originate from band crossings in a three-dimensional system, FeSe can be an intriguing platform to study such unique quantum states, which have recently been discussed in the Ba(FeAs)$_2$, Na$_3$Bi and Cd$_3$As$_2$ \cite{24, 29, 30, 31, 32}.

\section{Conclusions}
We investigated the electronic structures of FeSe, using a high quality single crystal synthesized by a recently reported new method, in the framework of a semiclassical transport theory.
The mobility spectrum employed as a powerful analytical method clearly demonstrated that the electronic structure in the low temperature orthogonal phase can be described as an almost compensated semimetal involving a minority band with ultrafast carrier mobility.
The ultrafast electron-like carriers could be interpreted as originating either from the Dirac cone or the large anisotropy of FS's.
Moreover, a remarkable reduction in carrier number and an enhancement in carrier mobility were simultaneously observed below 120 K, which is higher than $T^{\ast}$.
This significant change in the electronic structure was reasonably understood in terms of the development of electronically driven nematic ordering, being in good agreement with the recent ARPES experiments and their interpretations \cite{26}.

\section*{Acknowledgements}
We are grateful to Quynh T. N. Phan to support constructions of electrical transport measurement system under high magnetic fields and for K. Saito (Common Equipment Unit of Advanced Institute for Materials Research, Tohoku University) to support the XRD measuremetns
The research was partially supported by Grant-in-Aid for Young Scientists (B) (23740251) and the Joint Studies Program (2013) of the Institute for Molecular Science .



\clearpage

\begin{figure}[h]
\includegraphics[width=0.95\linewidth]{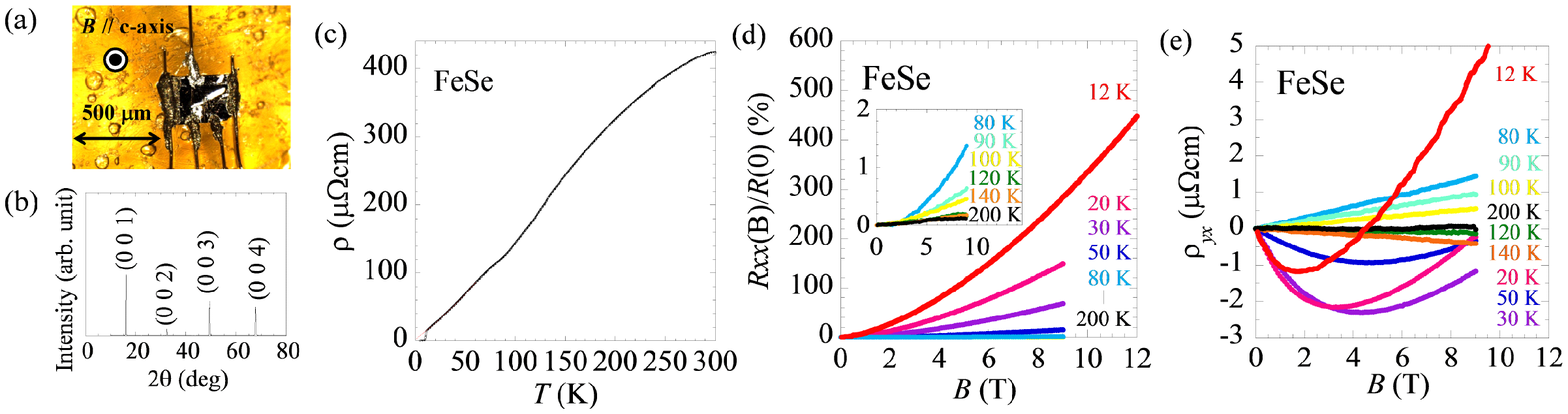}
\caption{(color online)(a) Schematic picture of the transport measurements of FeSe under magnetic fields. (b) X-ray diffraction profile of the FeSe single crystal along the c-axis direction at room temperature. (c) Temperature dependence of the resistivity of FeSe single crystal. (d), (e) Temperature evolution of electronic structures of FeSe and magnetic field dependence of magnetoresistance ($R_{xx}$($B$)/$R$(0) ) and Hall resistivity ($\rho_{yx}$) at various temperatures between 12 and 200 K. Inset of (b) shows a magnified plot of $R_{xx}$($B$)/$R$(0) between 80 - 200 K.}
\end{figure}

\clearpage

\begin{figure}[h]
\includegraphics[width=0.5\linewidth]{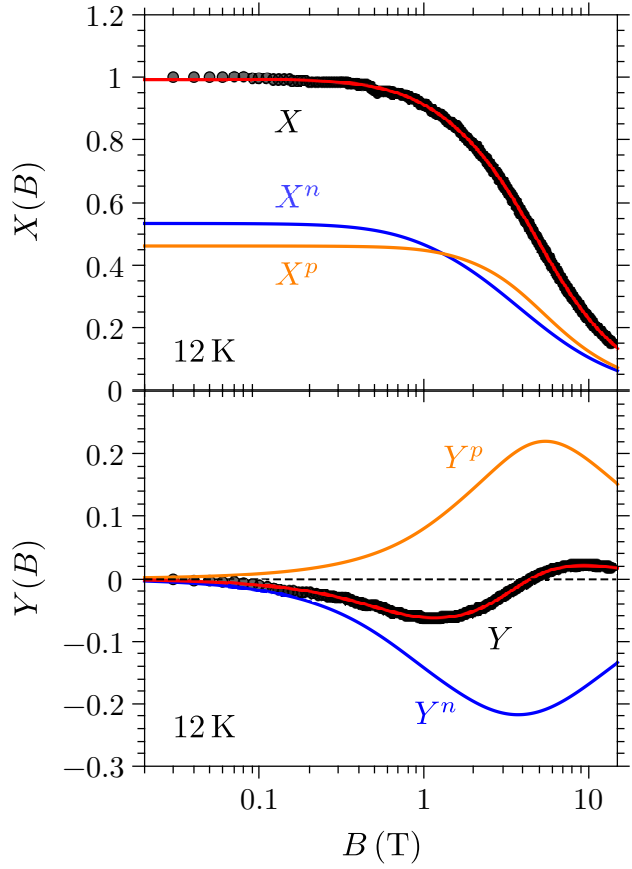}
\caption{(color online) (a), (b) The normalized conductivities $X(B)$ and $Y(B)$ calculated from the experimental data. Solid lines are patial conductivities of electron-like (blue), hole-like (orange) carriers, $[X^{(n)}(B),\,Y^{(n)}(B)]$ and $[X^{(p)}(B),\,Y^{(p)}(B)]$ employing the KK transformation. Also the summation of $[X^{(n)}(B),\,Y^{(n)}(B)]$ and $[X^{(p)}(B),\,Y^{(p)}(B)]$ ($[X(B), Y(B)]$) was plotted in the solid lines (red).}
\end{figure}

\begin{figure}[h]
\includegraphics[width=0.8\linewidth]{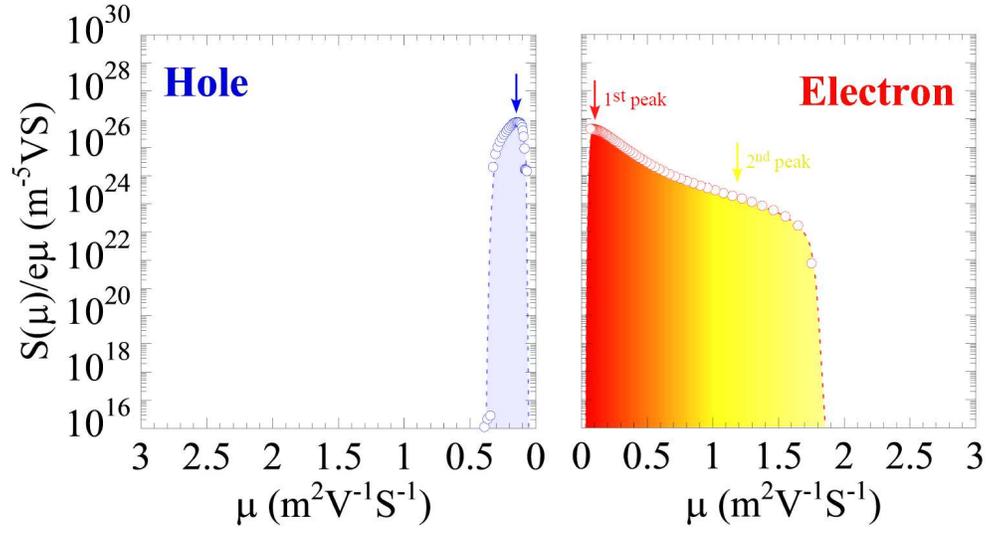}
\caption{(color online) Mobility spectra of electron-like and hole-like carriers for FeSe in the low temperature orthogonal phase displayed on a semi-logarithmic scale.}
\end{figure}

\clearpage

\begin{figure}[h]
\includegraphics[width=0.8\linewidth]{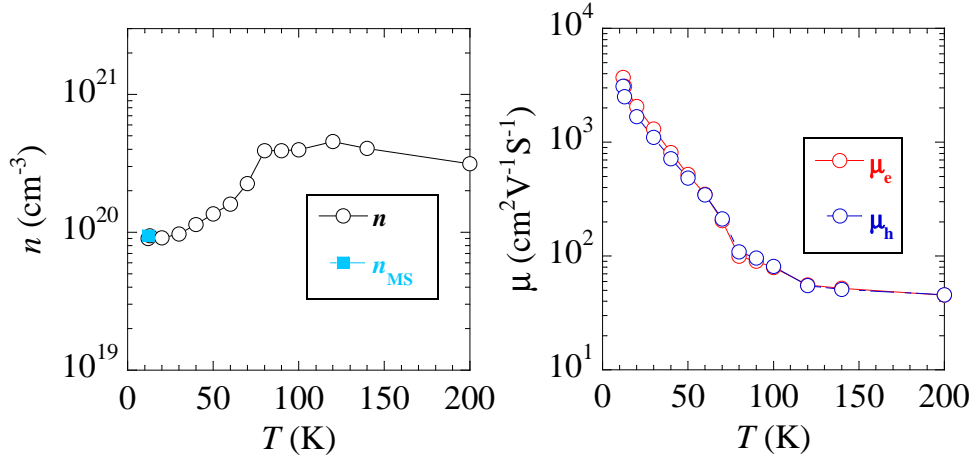}
\caption{(color online) (a), (b) Temperature dependences of carrier numbers ($n$) and electron (hole) mobility $\mu_e$ ($\mu_h$) derived using a two carrier-type semiclassical model in the low $B$ limit, assuming an  electron and hole compensated electronic structure. $n_{MS}$ is the averaged carrier number of electron-like and hole-like carriers ($n_{MS}$$^k$,$k$ = $n$, $p$) estimated from the mobility spectra. The error bars of $n$, $\mu_e$, and $\mu_h$ are estimated from errors of least square fit of transport parameters and are smaller than size of symbols.}
\end{figure}

\end{document}